\documentstyle[prl,aps,preprint]{revtex}
\begin{document}
\draft
\title{
\begin{flushright}
{\normalsize {\sf TITCMT-95-28}}
\end{flushright}
{\bf Quantum Hopfield Model}}
\author{Yoshihiko Nonomura\cite{nono} and Hidetoshi Nishimori}
\address{Department of Physics, Tokyo Institute of Technology,\\
         Oh-okayama, Meguro-ku, Tokyo 152, Japan}
\date{\today}
\maketitle
\begin{abstract}
The Hopfield model in a transverse field is investigated
in order to clarify how quantum fluctuations affect the
macroscopic behavior of neural networks. Using the Trotter
decomposition and the replica method, we find that the $\alpha$
(the ratio of the number of stored patterns to the system size)-$\Delta$
(the strength of the transverse field) phase diagram of this model
in the ground state resembles the $\alpha$-$T$ phase diagram of
the Hopfield model quantitatively, within the replica-symmetric and
static approximations. This fact suggests that quantum fluctuations
play quite similar roles to thermal fluctuations in neural networks
as long as macroscopic properties are concerned.\\
\end{abstract}
\pacs{PACS numbers: 05.50.+q, 87.10.+e, 75.10.Jm, 75.10.Nr}
\narrowtext
Neural networks have been investigated very actively in
the context of physics and engineering in terms of simple
mathematical models inspired by  anatomical and physiological
facts about the brain\cite{hertz,amit}.
In these models, the state of a neuron is often described
by an Ising spin, corresponding to the firing or at-rest state.
Neurons are connected with each other by long-range interactions,
and if these interactions are chosen suitably, some
fixed patterns of spins remain dynamically stable.
Thus, the system works as an associative memory.

In the time-evolution process of real neurons, randomness
appears in signal transmission at a synapse: A pulse
reaching the terminal bulb of an axon does not always
result in the release of neuro-transmitters contained in vesicles.
Usually, this randomness in signal transmission has been
taken into account in models as thermal fluctuations, which leads
to a statistical-mechanical formulation of neural networks
(see Sec. 2.1.3. of Ref.~\cite{amit}).
That is, the ``Hamiltonian'' ${\cal H}$ is defined so as to give stable
fixed patterns of the relevant network as global or local minima
of the energy landscape.
The ``temperature'' $T$ is next introduced and the ``partition
function" is defined by $Z\equiv {\rm Tr}e^{-{\cal H}/T}$ for a
given sample of embedded patterns.  The ``free energy" is then
obtained from the quenched average over the
samples as $F=-T\langle\langle\log Z\rangle\rangle$.

However, detailed considerations of the origin of randomness
in signal transmission
suggest that quantum effects may be an important driving force to cause
uncertainty in the release of neuro-transmitters from vesicles into
the synaptic cleft. For example, Stapp pointed out\cite{symqm}
that stochastic characters of signal transmission at synapses
may be explained by quantum uncertainty in the positions of
calcium ions during migration in the terminal bulb of an axon.
Beck and Eccles argued\cite{symqm2} that quantum fluctuations can be
of comparable order as thermal fluctuations in the hydrogen bridges
within axon terminals which control the exocytosis of synaptic vesicles.
These investigations strongly indicate the necessity to treat
randomness in the signal transmission in terms of quantum mechanics.

Under these motivations, we investigate a neural network
with quantum fluctuations.  In the conventional
statistical-mechanical approach to neural networks\cite{hertz,amit},
the parameter (temperature) $T$ is introduced
into a  Hamiltonian system.  This parameter does not necessarily reflect
directly the detailed stochastic properties of the original
 time-dependent neuron system.  Nevertheless, the effect
of the parameter $T$ is still regarded as a prototype of thermal
fluctuations. Similarly, the introduction of quantum fluctuations at
the level of the Hamiltonian formulation of the problem is expected
to be helpful in clarifying the roles of quantum effects in
the original system.  Admittedly, the model defined below
is not a faithful reproduction of real processes in the brain.
However, our purpose is not to explain the brain itself 
in detail\cite{comment}. We rather aim to clarify the 
statistical-mechanical roles of quantum fluctuations 
inlarge-scale networks at a phenomenological level. 
We believe that our model serves as a first step toward 
this goal.  Another motivation to develop the following
argument is that our method of investigation provides
a typical framework to treat quantum spin systems with
quenched randomness.

Let us therefore consider the Hopfield model\cite{hop}
in a transverse field,
\begin{equation}
  \label{qhopham}
  {\cal H}=-\sum_{i,j\atop i\neq j}J_{ij}\sigma_{i}^{z}\sigma_{j}^{z}
           -\Delta\sum_{i}\sigma_{i}^{x}
           \equiv{\cal H}_{0}+{\cal H}_{1}\ ,
\end{equation}
with the synaptic weight $J_{ij}$ defined by the Hebb rule,
\begin{equation}
  \label{weight}
  J_{ij}=\frac{1}{N}\sum_{\mu=1}^{p}\xi_{i}^{\mu}\xi_{j}^{\mu}\ .
\end{equation}
The transverse field triggers quantum tunneling from one state to another.
Even when $T$ is vanishing, phase transitions are expected to occur
as the strength of the transverse field $\Delta$ is varied.

Since ${\cal H}_{0}$ and ${\cal H}_{1}$ do not commute with
each other, we use the Trotter decomposition\cite{stdec} as
\begin{equation}
  \label{decomposition}
  Z={\rm Tr}e^{-\beta{\cal H}}
   =\lim_{M\to\infty}{\rm Tr}\left(e^{-\beta{\cal H}_{0}/M}
                                   e^{-\beta{\cal H}_{1}/M}
                                   \right)^{M}\ ,
\end{equation}
with $\beta\equiv 1/T$. Although the parameter $T$ seems 
indispensable in the present formulation, the existence 
of the $T\to 0$ limit can be justified\cite{stproof}, 
at least for the present type of quantum fluctuations. 
When the number of embedded patterns $p$ becomes infinite 
($\alpha\equiv p/N\sim O(1)$), the self-averaging property 
does not hold any more\cite{ags}, and the random average 
with respect to samples should be taken 
on the basis of the replica method,
\begin{equation}
  \label{decen}
  f\equiv-\frac{1}{N\beta}\langle\langle \log Z\rangle\rangle
        =-\frac{1}{N\beta}\lim_{n\to 0}
          \frac{\langle\langle Z^{n}\rangle\rangle-1}{n}\ .
\end{equation}
Actual evaluation of the above expression can be carried out
by applying the method of Amit {\it et al.}\cite{ags} to the
decomposed system (\ref{decomposition}). After long but straightforward
calculations\cite{qnnlong}, the free energy per spin is expressed
using only the Ising variable $\sigma_{\rho}(k)=\pm 1$ as
\begin{eqnarray}
  \label{exactf}
  nf&=&\frac{1}{2M}\sum_{k=1}^{M}\sum_{\mu=1}^{p}\sum_{\rho=1}^{n}
                   \left(m_{\rho}^{\mu}(k)\right)^{2}
      +\frac{\alpha}{2\beta}\sum_{k=1}^{M}\sum_{\rho=1}^{n}
                            \log\lambda_{\rho}(k)\nonumber\\*
    &+&\frac{\alpha\beta}{2M^{2}}\sum_{k=1}^{M}\sum_{l=1}^{M}
                                 \sum_{\rho=1}^{n}\sum_{\sigma=1,(\ne\rho)}^{n}
                                 r_{\rho\sigma}(k,l)q_{\rho\sigma}(k,l)
      +\frac{\alpha\beta}{2M^{2}}\sum_{k=1}^{M}\sum_{l=1}^{M}\sum_{\rho=1}^{n}
                                 t_{\rho}(k,l)S_{\rho}(k,l)\nonumber\\*
    &-&T\Biggl\langle\Bigg\langle\log\sum_{\sigma}
        \exp\Biggl(\frac{\beta}{M}\sum_{k,\mu,\rho}
                   m_{\rho}^{\mu}(k)\xi^{\mu}\sigma_{\rho}(k)
                   -\frac{1}{2}\log\tanh\frac{\beta\Delta}{M}
                    \sum_{k,\rho}\sigma_{\rho}(k)\sigma_{\rho}(k+1)\nonumber\\*
    &&\hspace{3cm} +\frac{\alpha\beta^{2}}{2M^{2}}
                    \sum_{k,l}\sum_{\rho,\sigma\atop\rho\neq\sigma}
                    r_{\rho\sigma}(k,l)\sigma_{\rho}(k)\sigma_{\sigma}(l)
                   +\frac{\alpha\beta^{2}}{2M^{2}}
                    \sum_{k,l}\sum_{\rho}
                    t_{\rho}(k,l)\sigma_{\rho}(k)\sigma_{\rho}(l)
                   \Biggr)\Bigg\rangle\Biggr\rangle\ ,
\end{eqnarray}
where $\{\lambda_{\rho}(k)\}$ stand for the eigenvalues
of the following $Mn\times Mn$ matrix,
\begin{equation}
  \label{genmat}
  \Lambda_{k\rho,l\sigma}=\delta_{kl}\delta_{\rho\sigma}
                         -\frac{\beta}{M}q_{\rho\sigma}(k,l)
                         -\frac{\beta}{M}\delta_{\rho\sigma}S_{\rho}(k,l)\ ,
\end{equation}
with
\begin{eqnarray}
  q_{\rho\sigma}(k,l)
  &=&\frac{1}{N}\sum_{i=1}^{N}\sigma_{i\rho}(k)\sigma_{i\sigma}(l)\ ,\\
  S_{\rho}(k,l)
  &=&\frac{1}{N}\sum_{i=1}^{N}\sigma_{i\rho}(k)\sigma_{i\rho}(l)\ .
\end{eqnarray}

There appear five kinds of order parameters $m_{\rho}^{\mu}(k)$,
$q_{\rho\sigma}(k,l)$, $r_{\rho\sigma}(k,l)$, $S_{\rho}(k,l)$, and
$t_{\rho}(k,l)$. The variables $\rho$ and $\sigma$ represent replica
indices, and $k$ and $l$ stand for Trotter indices. From the saddle-point
equations with respect to these order parameters, they are expressed as
\begin{eqnarray}
  m_{\rho}^{\mu}(k)
  &=&\left\langle\left\langle\frac{1}{N}\sum_{i=1}^{N}
     \xi_{i}^{\mu}\langle\sigma_{i\rho}(k)\rangle
     \right\rangle\right\rangle\ ,\\
  q_{\rho\sigma}(k,l)
  &=&\left\langle\left\langle\frac{1}{N}\sum_{i=1}^{N}
     \langle\sigma_{i\rho}(k)\rangle\langle\sigma_{i\sigma}(l)\rangle
     \right\rangle\right\rangle\ ,\\
  r_{\rho\sigma}(k,l)
  &=&\frac{1}{\alpha}\sum_{\mu=s+1}^{p}
     \left\langle\left\langle m_{\rho}^{\mu}(k)m_{\sigma}^{\mu}(l)
                             \right\rangle\right\rangle\ ,\\
  S_{\rho}(k,l)
  &=&\left\langle\left\langle\frac{1}{N}\sum_{i=1}^{N}
     \langle\sigma_{i\rho}(k)\rangle\langle\sigma_{i\rho}(l)\rangle
     \right\rangle\right\rangle\ ,\\
  t_{\rho\sigma}(k,l)
  &=&\frac{1}{\alpha}\sum_{\mu=s+1}^{p}
     \left\langle\left\langle m_{\rho}^{\mu}(k)m_{\rho}^{\mu}(l)
                             \right\rangle\right\rangle\ ,
\end{eqnarray}
where $s$ denotes the number of condensed patterns for which the
parameter $m_{\rho}^{\mu}(k)$ remains finite in the $N\to\infty$
limit. Especially, in the single-retrieval case, $s=1$.
Although the first three order parameters also appear
in the replica calculation of the Hopfield model\cite{ags}
aside from the dependence on Trotter indices, the last two order
parameters are specific to the present model. In the present
type of path-integral formulations of quantum spin systems,
the Trotter index $k$ can be identified with an imaginary-time
index. Then, the quantities $S_{\rho}(k,l)$ and $t_{\rho}(k,l)$
may be considered to be related with dynamical properties
of the quantum spin system described by the Hamiltonian
(\ref{qhopham}), though this ``dynamics" is nothing to
do with the real time evolution of original neurons.

For further analytic calculations, we introduce the
replica-symmetric approximation\cite{ags,qsk} defined by
$m_{\rho}^{\mu}(k)\to m^{\mu}(k)$, $q_{\rho\sigma}(k,l)\to q(k,l)$
for $\rho\neq\sigma$, $r_{\rho\sigma}(k,l)\to r(k,l)$,
$S_{\rho}(k,l)\to S(k,l)$ and $t_{\rho}(k,l)\to t(k,l)$.
As will be shown later, this approximation is justified in most of the
parameter regions of $\alpha$ and $\Delta$ even at $T=0$. The resulting
expression of the free energy is still too complicated for analytic
studies. Thus, we adopt the static approximation\cite{qsk} in which
we set $m^{\mu}(k)\to m^{\mu}$, $q(k,l)\to q$, $r(k,l)\to r$,
$t(k,l)\to t$, $S(k,l)\to S$ for $k\neq l$ and $S(k,l)\to 1$ for $k=l$.
This name (static approximation) originates from the fact that the
correlations along the Trotter (or {\it imaginary-time}) direction are
averaged out in this approximation. Among these five order parameters,
the static approximation seems quite reasonable for $m^{\mu}(k)$,
$q(k,l)$ and $r(k,l)$, because the physical meaning of these order
parameters (overlaps of an embedded pattern, SG order parameters and
effects of uncondensed patterns) suggests very slight dependence on
Trotter indices. On the other hand, the static approximation leads to
a superficial inconsistency for $S(k,l)$ and $t(k,l)$.
That is, direct calculations of these
quantities within the static approximation\cite{qnnlong} result in
the Trotter-layer dependence of $S(k,l)$ and $t(k,l)$. However,
these quantities appear in the free energy only as the sum over
all the Trotter layers. Thus, the static approximation recovers
consistency after such a summation is carried out\cite{qnnlong}.

On the basis of these approximations, the $Mn\times Mn$ matrix
(\ref{genmat}) can be diagonalized for any values of $M$ and $n$.
Taking the limits $M\to\infty$ and $n\to 0$, we have\cite{qnnlong}
\begin{eqnarray}
  \label{finf}
  f&=&\frac{1}{2}\sum_{\mu}(m^{\mu})^{2}
     +\frac{\alpha}{2\beta}\left[\log(1+\beta q-\beta S)
                                 -\frac{\beta q}{1+\beta q-\beta S}
                                 -\beta(1-S)\right]
     -\frac{1}{2}\alpha\beta(rq-ts)\nonumber\\*
   &-&T\left\langle\left\langle\int_{-\infty}^{+\infty}Dz
            \log\left[2\int_{-\infty}^{+\infty}
                      {\textstyle Dw\cosh\beta
                       \sqrt{\left(\sum_{\mu}\xi^{\mu}m^{\mu}+\sqrt{\alpha r}z
                                   +\sqrt{\alpha(t-r)}w\right)^{2}+\Delta^{2}}
                      }\right]\right\rangle\right\rangle\ ,
\end{eqnarray}
where $Dx$ denotes the Gaussian measure, $dx e^{-x^{2}/2}/\sqrt{2\pi}$.
When we take the $\Delta\to 0$ limit, this expression coincides
exactly with that of the Hopfield model\cite{ags}, as expected.
Extremizing this free energy with respect to $m$, $q$, $r$, $S$ and $t$,
we obtain a set of equations which describes static
properties of the present model for any $\alpha$, $\Delta$ and $T$.

The limitation of the static approximation
has already been recognized in the SK model in a
transverse field\cite{qsk}. In the SK model, this approximation
overestimates the critical transverse field $\Delta_{{\rm c}}$, because the
symmetry between Trotter layers is broken owing to the infinite degeneracy
of ground states in each layer. This symmetry breaking is missed by the
static approximation. In spite of such difficulties, this approximation
holds at least in the vicinity of $\Delta=0$ or $\alpha=0$. In the former case
($\Delta\approx 0$), quantum fluctuations are small, and therefore
the validity of the static approximation is easily accepted.
In the latter case ($\alpha\approx 0$), the free energy of
the model (\ref{qhopham}) can be calculated exactly without
using the replica method for $p$ finite (or $\alpha\to 0$), and the 
result is consistent with the present one\cite{qnnlong}. From a
physical point of view, the finite-$p$ case is a straightforward
generalization of the Mattis model\cite{mattis}, and the symmetry
between Trotter layers would not be broken.

In order to see the effects of quantum fluctuations in the
absence of thermal noise, we now consider the case of $T=0$.
Then, the phase diagram within the replica-symmetric and static
approximations can be obtained from the following set of equations,
\begin{eqnarray}
  \label{feq1}
  m&=&\int_{-\infty}^{+\infty}Dz\frac{m+\sqrt{\alpha r}z}
           {\sqrt{\left(m+\sqrt{\alpha r}z\right)^{2}+\Delta^{2}}}\ ,\\
  \label{feq2}
  q&=&\int_{-\infty}^{+\infty}Dz\frac{\left(m+\sqrt{\alpha r}z\right)^{2}}
           {\left(m+\sqrt{\alpha r}z\right)^{2}+\Delta^{2}}\ ,\\
  C&=&\int_{-\infty}^{+\infty}Dz\frac{\Delta^{2}}
           {\left[\left(m+\sqrt{\alpha r}z\right)^{2}+\Delta^{2}
                  \right]^{3/2}}\ ,\\
  r&=&\frac{q}{(1-C)^{2}}\ .
  \label{feq3}
\end{eqnarray}
The parameters $S$ and $t$ in Eq.\ (\ref{finf}) satisfy $S=q$ and $t=r$
(the differences $S-q$ and $t-r$ are of the order of $T$), respectively.
Note that the new parameter $C$ originates from this $O(T)$ contribution.

The solution of Eqs.\ (\ref{feq1})--(\ref{feq3}) reveals that
the structure of the phase diagram is quite similar to that
of the Hopfield model, as shown in Fig.\ \ref{phasefig}.
The phase boundary between the retrieval phase ($m\neq 0$ and $q\neq 0$)
and the SG phase ($m=0$ and $q\neq 0$) can be evaluated only by solving
numerically the integral equations (\ref{feq1})--(\ref{feq3}) directly,
because this transition is of first order.  Unexpectedly,
the shape of this phase
boundary is quite similar to that of the Hopfield model\cite{ags}
even quantitatively, when the parameter $T$ is replaced by $\Delta$.
The asymptotic form of the phase boundary around $\alpha=0$ turns out
to be $\Delta_{{\rm c}}\simeq 1-1.95\sqrt{\alpha}$\cite{qnnlong}
in exact agreement with that of the Hopfield model,
$T_{{\rm c}}\simeq 1-1.95\sqrt{\alpha}$\cite{ags}.
The critical capacity $\alpha_{{\rm c}}\simeq 0.1379$ at $\Delta=0$
agrees with the corresponding Hopfield value\cite{ags} by definition.
The line of the second-order phase transition between the SG
phase and the paramagnetic phase ($m=0$ and $q=0$) can be calculated
analytically by setting $m=0$ and expanding the formulas
(\ref{feq2})--(\ref{feq3}) to the lowest order of $r$ ($r$ and $q$
are of the same order in the vicinity of this phase boundary).
The critical value of the transverse field is given by
$\Delta_{{\rm sg}}=1+\sqrt{\alpha}$.
This expression is {\it exactly} the same as that of the
Hopfield model\cite{ags}, when $\Delta$ is replaced by $T$,
again an unexpected result.

The results presented above can be derived much more easily.
First, we apply the mean-field approximation\cite{nnges}
to the Hamiltonian (\ref{qhopham}):
\begin{equation}
  \label{effham}
  {\cal H}^{{\rm eff}}
 =-\frac{1}{N}\sum_{i=1}^{N}\sum_{\mu=1}^{p}\xi_{i}^{\mu}\xi_{j}^{\mu}
              \langle\sigma_{i}^{z}\rangle\sigma_{j}^{z}
  -\Delta\sigma_{j}^{x}\ .
\end{equation}
Next, we require the self-consistency condition as
$\langle\sigma_{i}^{z}\rangle=\langle\sigma_{j}^{z}\rangle_{{\rm eff}}$,
where $\langle\cdots\rangle_{{\rm eff}}$ stands for the average with
respect to the effective Hamiltonian (\ref{effham}), and we have\cite{nnges}
\begin{equation}
  m_{\nu}\equiv\frac{1}{N}\sum_{i=1}^{N}\xi_{i}^{\nu}
                          \langle\sigma_{i}^{z}\rangle
              =\frac{1}{N}\sum_{i=1}^{N}\xi_{i}^{\nu}
               \frac{\sum_{\mu}\xi_{i}^{\mu}m_{\mu}}
                    {\sqrt{\left(\sum_{\mu}\xi_{i}^{\mu}m_{\mu}
                                 \right)^{2}+\Delta^{2}}}
               {\textstyle
                \tanh\beta\sqrt{\left(\sum_{\mu}\xi_{i}^{\mu}m_{\mu}
                                            \right)^{2}+\Delta^{2}}
               }\ .
\end{equation}
For finite $p$, all the results of the replica calculations\cite{qnnlong}
can be obtained from this equation. Even if $p$ is of the order of $N$,
Geszti's approximation\cite{nnges} can be applied to this model to obtain
the same set of equations given in (\ref{feq1})--(\ref{feq3})\cite{qnnlong}.
This fact supports the validity of the replica-symmetric
and static approximations used above.

Since the expression of the free energy (\ref{finf}) has been
derived explicitly, we can go further than the above mean-field
approach. Comparing the values of the free energy of the retrieval
and SG solutions, we obtain the phase boundary between the
global-minimum retrieval phase (R-I) and the local-minimum
retrieval phase (R-II) as displayed in Fig.\ \ref{phasefig}.
This phase boundary is also similar to
that of the Hopfield model in the vicinity of $\Delta=1$:
$\Delta_{{\rm c}}=1-2.0625\sqrt{\alpha}$, while in the Hopfield model,
$T_{{\rm c}}\simeq 1-2.6\sqrt{\alpha}$\cite{ags}. Near $\Delta=0$,
reentrant behavior is observed. Although this behavior has
not been reported explicitly in the Hopfield model\cite{ags},
we have found by detailed calculations that this reentrant
behavior also takes place in the classical case. These classical
and quantum phase boundaries resemble each other
even in this region.

The AT line can be calculated by generalizing the analysis in
the case of the Hopfield model\cite{ags}. After some tedious
calculations\cite{qnnlong}, we find that this line is given
by the solution of the following equation,
\begin{equation}
  q=\alpha r\Delta^{4}\int_{-\infty}^{+\infty}Dz
    \frac{1}{\left[\left(m+\sqrt{\alpha r}z\right)^{2}+\Delta^{2}
                   \right]^{3}}\ .
\end{equation}
The AT line is shown in Fig.\ \ref{phasefig}. A nontrivial
reentrant phase transition between the SG and retrieval phases also
seems to occur in the vicinity of $\Delta=0$ and $\alpha=0.1379$.
However, the reentrant region around $\Delta=0$ and $\alpha=0.1379$
lies below the AT line. A similar property is reported in a recent
detailed calculation of the Hopfield model\cite{hopdet}. The shape
of the AT line of the present quantum model is quite similar to
that of the Hopfield model given in Ref.\ \cite{hopdet},
if the parameter $\Delta$ is replaced by $T$.

In summary, the Hopfield model in a transverse field was introduced 
and investigated using the Trotter decomposition and the replica method. 
The replica-symmetric and static approximations enabled us to treat the 
problem analytically.  We found that the ground-state phase diagram 
of this model is quite similar to that of the Hopfield model, when the 
strength of the transverse field $\Delta$ is replaced by the temperature 
$T$. We also showed that the same phase diagram can be obtained by the 
mean-field approximation. Using the expression of the free energy, 
the phase boundary on which the retrieval states become the global 
minima and the AT line have been obtained. These two boundaries 
are also similar to those of the Hopfield model.

It is appropriate to stress here that the qualitative and quantitative 
coincidence of the properties of the $T=0$ quantum system with 
those of the finite-temperature classical system is quite 
nontrivial.  In consideration of quantum aspects of 
randomness in signal transmission\cite{symqm,symqm2}, 
our results may serve as an {\it a posteriori} justification 
of the conventional classification of macroscopic phases of 
the Hopfield model\cite{hertz,amit} if this model is regarded 
as a primitive but first step toward understanding emergent 
properties of the brain functioning. The method in this 
Letter provides also a statistical-mechanical framework 
for analysis of systems in which both quenched randomness 
and quantum fluctuations are present. 

One of the present authors (Y.\ N.) is grateful for the financial support 
of the Japan Society for the Promotion of Science for Japanese Junior 
Scientists. Numerical calculations were performed on FACOM VPP 500 
at the Institute for Solid State Physics, University of Tokyo. 
\noindent
\begin{figure}[h]
\caption{
The ground-state phase diagram of the Hopfield model in a 
transverse field. ``R-I" stands for the retrieval phase in which 
the retrieval states are the global minima, and ``R-II" denotes the 
retrieval phase where the retrieval states are the local minima. 
The dashed line represents the AT line. The vicinity of
$\alpha=0.1379$ and $\Delta=0$ is shown magnified in the inset. 
}
\label{phasefig}
\end{figure}

\begin{references}
\bibitem{comment}
One may argue in this regard that the Hopfield model without the
transverse-field term is already inadequate as a realistic model
of the brain.  The Hopfield model has nevertheless been the target of
active research.
\end{references}
\end{document}